\pdfoutput=1
\documentclass[sigconf]{acmart}
\usepackage{listings}
\usepackage{enumitem}
\definecolor{codegreen}{rgb}{0,0.6,0}
\definecolor{codegray}{rgb}{0.5,0.5,0.5}
\definecolor{codered}{rgb}{0.6,0,0}
\definecolor{backcolour}{rgb}{0.95,0.95,0.95}
\lstdefinestyle{mystyle}{
    backgroundcolor=\color{backcolour},   
    commentstyle=\color{codegreen},
    keywordstyle=\color{magenta},
    numberstyle=\tiny\color{codegray},
    stringstyle=\color{codered},
    basicstyle=\ttfamily\scriptsize,
    breakatwhitespace=false,         
    breaklines=true,                 
    captionpos=b,                    
    keepspaces=true,                 
    numbers=left,                    
    numbersep=5pt,                  
    showspaces=false,                
    showstringspaces=false,
    showtabs=false,                  
    tabsize=2
}
\usepackage{array}

\usepackage{tikz}
\newcommand*\circled[1]{\tikz[baseline=(char.base)]{
            \node[shape=circle,draw,inner sep=0.5pt] (char) {#1};}}

\AtBeginDocument{%
  }

\copyrightyear{2025} 
\acmYear{2025} 
\setcopyright{rightsretained}
\acmConference[ASIA CCS '25]{ACM Asia Conference on Computer and Communications Security}{August 25--29, 2025}{Hanoi, Vietnam}
\acmBooktitle{ACM Asia Conference on Computer and Communications Security (ASIA CCS '25), August 25--29, 2025, Hanoi, Vietnam}\acmDOI{10.1145/3708821.3710823}
\acmISBN{979-8-4007-1410-8/25/08}

\usepackage{hyperref}
\hypersetup{unicode}

\begin{document}

\title[Ruling the Unruly]{Ruling the Unruly: Designing Effective, Low-Noise Network Intrusion Detection Rules for Security Operations Centers}

% \author{anonymous author}
% \email{anonymous@example.com}
% \orcid{0000-0000-0000-0000}
% \affiliation{%
%   \institution{Institution}
%   \city{City}
%   \country{Country}
% }
% ORCID of the author: http://orcid.org/0000-0002-6490-4768
\author{Koen T. W. Teuwen}
\email{k.t.w.teuwen@tue.nl}
\orcid{0000-0002-6490-4768}
\affiliation{%
  \institution{Eindhoven University of Technology}
  \city{Eindhoven}
  \country{The Netherlands}
}

\author{Tom Mulders}
\email{t.r.j.mulders@tue.nl}
\orcid{0009-0004-4387-3458}
\affiliation{%
  \institution{Eindhoven University of Technology}
  \city{Eindhoven}
  \country{The Netherlands}
}

\author{Emmanuele Zambon}
\email{e.zambon.n.mazzocato@tue.nl}
\orcid{0000-0002-8079-4087}
\affiliation{%
  \institution{Eindhoven University of Technology}
  \city{Eindhoven}
  \country{The Netherlands}
}

\author{Luca Allodi}
\email{l.allodi@tue.nl}
\orcid{0000-0003-1600-0868}
\affiliation{%
  \institution{Eindhoven University of Technology}
  \city{Eindhoven}
  \country{The Netherlands}
}

% \renewcommand{\shortauthors}{K.T.W. Teuwen et al.}

\begin{abstract}
\looseness=-1 Many Security Operations Centers (SOCs) today still heavily rely on signature-based Network Intrusion Detection Systems (NIDS) such as Suricata. The specificity of intrusion detection rules and the coverage provided by rulesets are common concerns within the professional community surrounding SOCs, which impact the effectiveness of automated alert post-processing approaches. We postulate a better understanding of factors influencing the quality of rules can help address current SOC issues.
In this paper, we characterize the rules in use at a collaborating commercial (managed) SOC serving customers in sectors including education and IT management.
During this process, we discover six relevant design principles, which we consolidate through interviews with experienced rule designers at the SOC. 
We then validate our design principles by quantitatively assessing their effect on rule specificity.
We find that several of these design considerations significantly impact unnecessary workload caused by rules.
For instance, rules that leverage proxies for detection, and rules that do not employ alert throttling or do not distinguish (un)successful malicious actions, cause significantly more workload for SOC analysts. Moreover, rules that match a generalized characteristic to detect malicious behavior, which is believed to increase coverage, also significantly increase workload, suggesting a tradeoff must be struck between rule specificity and coverage.
We show that these design principles can be applied successfully at a SOC to reduce workload whilst maintaining coverage despite the prevalence of violations of the principles.
\end{abstract}

\begin{CCSXML}
<ccs2012>
<concept>
<concept_id>10002978.10002997.10002999</concept_id>
<concept_desc>Security and privacy~Intrusion detection systems</concept_desc>
<concept_significance>500</concept_significance>
</concept>
<concept>
<concept_id>10002978.10003029.10011703</concept_id>
<concept_desc>Security and privacy~Usability in security and privacy</concept_desc>
<concept_significance>500</concept_significance>
</concept>
<concept>
<concept_id>10002978.10003014</concept_id>
<concept_desc>Security and privacy~Network security</concept_desc>
<concept_significance>300</concept_significance>
</concept>
</ccs2012>
\end{CCSXML}

% \ccsdesc[500]{Security and privacy~Usability in security and privacy}
\ccsdesc[500]{Security and privacy~Intrusion detection systems}
\ccsdesc[300]{Security and privacy~Network security}

\keywords{Security Operations Center (SOC), Network Intrusion Detection System (NIDS), Network Intrusion Detection Rules}


% \pagestyle{plain}

\maketitle

\section{Introduction}
\label{sec:introduction}

Security Operations Centers (SOCs) are an important part of organizations' security strategies to establish defense-in-depth and increase their digital resiliency.
SOCs typically operate using a combination of automated detectors such as Host Intrusion Detection Systems (HIDS) or Network Intrusion Detection Systems (NIDS), followed by manual analysis of the security events collected from these tools. 
A NIDS commonly employed in SOCs is Suricata~\cite{Suricata, alert-alchemy, ruling-the-rules}, which is a rule-based intrusion detection system. 

\looseness=-1Modern SOCs suffer from several issues, and academic research is not always aligned with the problems SOC employees experience in practice \cite{Vielberth_Bohm_Fichtinger_Pernul_2020,Kokulu_Soneji_Bao_Shoshitaishvili_Zhao_Doupé_Ahn_2019}. For example, companies are unable to keep up with changes in the threat landscape due to a lack of coverage~\cite{Vielberth_Bohm_Fichtinger_Pernul_2020, Kokulu_Soneji_Bao_Shoshitaishvili_Zhao_Doupé_Ahn_2019}.
Other research work~\cite{ruling-the-rules, alert-alchemy, 99-false-positives} suggests a lack of \textit{specificity} is a bigger concern, causing analysts a tremendous amount of unnecessary workload.
To mitigate the workload imposed on SOC analysts by the volume of incoming alerts requiring investigation, SOCs can employ various automated alert post-processing tools. These include alert classification \cite{Ede_Aghakhani_Spahn_Bortolameotti_Cova_Continella_Steen_Peter_Kruegel_Vigna_2022}, prioritization \cite{Vidovic_Tomicic_Slovenec_Mikuc_Brajdic_2021}, and aggregation \cite{Okutan_Yang_2019} methods, which are often based on Machine Learning (ML).
As pointed out by~\cite{alert-alchemy}, the attempts to use ML to improve SOC operations are ``end of pipe'' solutions designed to take large quantities of alerts as input in order to produce informative and actionable output that SOC analysts can use, but often fail to achieve this goal due to poor input quality. These approaches would greatly benefit from improving the quality of the alerts, and thus the rules generating them.

We postulate a better understanding of factors influencing the quality of rules can help address current SOC issues.
Although previous work has focused on characterizing rulesets and their quality~\cite{ruling-the-rules,alert-alchemy}, to the best of our knowledge, no characterization of the rules contained therein has been presented yet, and no actionable guideline has been built to facilitate the design of quality rules.
Even if a relation between rule coverage and specificity is known to exist~\cite{alert-alchemy}, no academic literature explores the relation between rule design principles and the coverage or specificity of the designed rules.
In this paper, we characterize the rules in use at a collaborating commercial (managed) SOC serving customers in sectors including education and IT management.
Specifically, we leverage the alerts generated by Suricata sensors deployed at two of the largest SOC customers, together with incident reports, to characterize the rules contributing to the detection of incidents (good rules) and the rules responsible for generating mainly ``noise'' (high-noise rules).
Despite originally assuming implementation-level characteristics (e.g., rule detection and metadata options) would be able to explain why certain rules are good and others are high-noise, we find that implementation-level features merely serve as a proxy for higher-level design characteristics.
During this process, we discover several relevant principles at the design level, which we consolidate through interviews with experienced rule designers at the SOC.
We then validate our design principles by quantitatively assessing their effect on rule specificity. 
We make the following contributions: 
\begin{itemize}[leftmargin=*]
    \item We present the first analysis of the quality of NIDS rules (as opposed to rulesets) deployed at a commercial SOC. The analysis is based on a ruleset comprising $290k$ unique rule revisions and on the $30M$ alerts generated by these rules during 11 months of deployment at two SOC customer networks.
    \item We find that rules leveraging atomic Indicators of Compromise (IoC) contribute the most to incident detection and a lack of coverage is not always considered a culprit. Recently introduced rules and legacy rules play an important role in providing coverage.
    \item A minority of rules is responsible for the majority of alerts, whereas a vast majority of rules never trigger. After examining high-noise rules, we found that they can be improved within the possibilities offered by the underlying detection technology.
    \item Based on our observations from the collected data, complemented by expert interviews, we formulate six design principles to improve rule specificity and coverage.
    \item Utilizing regression analysis, we find that several of the design principles have a significant impact on rule specificity. Interestingly, we find that one of the principles to improve coverage negatively affects specificity, suggesting a specificity and coverage must be balanced during rule design.
    \item We develop a tool\footnote{The code will be released as open source software under EUPL 1.2.} to identify rules deviating from the principles and find that a majority of the acquired (open and commercial) rules from the collaborating SOC likely uses proxies for detection, and fails to utilize exceptions and alert throttling.
\end{itemize}

The rest of this paper is organized as follows. We provide background information and discuss related work in Section~\ref{sec:background}. In Section~\ref{sec:methodology}, we discuss the methodology for our research. Thereafter, we present an exploration of the data we gathered, together with our rule quality findings in Section~\ref{sec:data_exploration}. 
In Section~\ref{sec:design-principles}, we present our rule design principles and analyze their effect on rule coverage and specificity. In Section~\ref{sec:operationalization}, we show how the design principles can be operationalized in the context of a SOC.
We discuss the implications of our results in Section~\ref{sec:discussion} along with the limitations of our empirical analyses. Section~\ref{sec:conclusion} concludes this work.

\section{Background and related work}
\label{sec:background}

\subsection{Security Operations \& Intrusion Detection}
\label{sec:background_soc}
\looseness=-1A \textit{Security Operations Center (SOC)} can be defined as a (part of) an organization that is responsible for monitoring assets to detect and respond to security incidents~\cite{Vielberth_Bohm_Fichtinger_Pernul_2020}. SOCs can be in-house or managed. Managed SOCs are external to the monitored organization and typically provide services to a number of other organizations. Typically, SOCs deploy network monitoring or host monitoring solutions to gain insight into their assets. Although SOCs may employ \textit{Host-based Intrusion Detection Systems (HIDS)}, we focus on \textit{Network-based Intrusion Detection Systems (NIDS)} in the remainder of this work.
To process the vast amount of monitored data streams, \textit{intrusion detection systems} are applied to detect undesirable behavior. Intrusion detection systems can be divided into misuse-based detection systems and anomaly detection systems \cite{liao-13}. \textit{Misuse-based} detection systems use knowledge of the characteristics of malicious behavior to perform detection, while \textit{anomaly-based} detection uses a baseline of normal behavior to detect deviations. Suricata is an example of a misuse-based NIDS and requires a \textit{ruleset} consisting of \textit{rules} implementing \textit{signatures} describing potential malicious behavior. In addition to the logs produced by these intrusion detection systems, more generic \textit{Security Monitoring Systems (NSM)} exist that can be used to describe the monitored data streams using key features and characteristics, such as Zeek \cite{Zeek}.
In addition, Tactics and Techniques from the MITRE taxonomy~\cite{mitre-taxonomy} are often used to distinguish between different actions that an attacker may perform. Rules may be designed to detect instances of specific ATT\&CK techniques.

In the remainder of this paper, \textit{alert data} refers to the logs generated by intrusion detection systems such as Suricata. \textit{Log data} refers to data describing other logs produced by tools not used to detect intrusions and, hence, also describes normal behavior. Analysts can leverage both alert and log data to manually detect security incidents with the support of data processing tools such as Security Information and Event Management tools (SIEM), which provide visualizations and can be used to correlate different logs. When the outcome of an alert investigation is escalated to a customer, we say an \textit{incident} has been detected. A SOC typically documents the incident in an incident report describing the nature of the incident. Even if not all incident reports may correspond to security incidents, they are necessary to trigger further investigation, such as OS integrity inspection, after non-conclusive investigations based on evidence extracted from network traffic only.

\subsection{Rulesets \& Rule Engineering}
\looseness=-1NIDS rules examined in this work are written according to the Suricata rule syntax~\cite{Suricata-docs}. Each rule begins with a \textit{header} specifying the \textit{action}, and \textit{protocol} and \textit{direction} of the inspected network packets. Each rule can have additional options within the \textit{body} of the rule. The most important options are the \textit{detection options}. They specify which buffer is matched against which string, bytes, or regular expression. \textit{Content modifiers} exist to change how and where the content is matched. A rule may combine positive and negative content matches and even specify \textit{distance} between content matches. Additionally, the \textit{flow} keyword enables restricting the inspection of packets sent only by servers or by clients. Notably, Suricata also includes options to set and check bits within and across connections using keywords such as \textit{flowbits} and \textit{xbits}, enabling stateful detection. Additionally, rules can use \textit{threshold} to limit how often and when they trigger. \textit{Non-detection options}, such as \textit{metadata} and \textit{classtype}, also exist to aid interpretation of generated alerts. Figure~\ref{fig:suricata-example}, contained in Appendix~\ref{sec:appendix_0}, shows an example of a rule and highlights the aforementioned different syntax features. The rule engineer may use arbitrary combinations of these syntax features to define rules as specific as possible (i.e., triggering only when a successful attack is present) while maintaining coverage of attack variants (i.e., avoiding overfitting to specific instances of that attack). Several broad categories of rules can be distinguished~\cite{et-category-descriptions}, including those leveraging Indicators of Compromise (IoCs), those leveraging information on known exploits, those leveraging information about vulnerable devices, and those leveraging information about rare deviations from common protocols. Rules can also be grouped according to the type of malicious action they detect or the considered protocol.

\looseness=-1The evolution of rules and rules, their raising of alerts, and their relation to incidents have been studied in the extant literature~\cite{ruling-the-rules}. Notable findings include that a minority of the rules is responsible for the majority of the alerts, only a fraction of the alerts are investigated, and most rules never receive updates on the detection option but only on metadata (which has no impact on triggering conditions). Previous research has also suggested that about one-fourth of all raised alerts correspond to unsuccessful attack attempts, and almost half correspond to \textit{benign triggers} (i.e., an alert that is fired based on the correct characteristic, but that can be explained by an acceptable behavior)~\cite{attack-attempts}. Building upon these works, our aim is to explain which rules are responsible for the vast majority of uninteresting alerts and which rule design principles play a role in that.

\looseness=-1Other researchers conducted interviews with professionals from various SOCs to better understand the workflows and decision-making of SOCs~\cite{alert-alchemy}. Additionally, they found that, while it is common to procure community and commercial rulesets, SOCs tune these rulesets by disabling rules. They also state that feedback loops to report False Positives (FPs) are not common practice. They found that rule specificity, especially, but also rule(set) coverage, are key considerations in SOCs, which must be balanced together. Our work aims to fill the gap on the factors that contribute to rule specificity.

Other research also reports on SOC processes following interviews and describes a SOC analyst who blamed a rule engineer for developing a high-noise rule that the analyst had to deal with~\cite{turning-contradictions}. A related study on FPs at SOCs concluded that most FPs correspond to benign behavior~\cite{99-false-positives}.
Regarding rule quality, the community surrounding Suricata has made several efforts to develop guidelines to promote rule quality, although these mostly revolve around the syntax and structure of rules~\cite{suricata-style-guide, snort-suricata-guide}.

\section{Methodology}
\label{sec:methodology}

\subsection{Data Provisioning}
\label{sec:data_sources}

\subsubsection{Monitored organizations}
For the analysis conducted in this paper, data was gathered from a managed SOC. The SOC monitors network interactions with approximately $100k-200k$ IP addresses, collectively responsible for approximately $1-10M$ connections per day. As part of their Service Level Agreement (SLA), each client organization has set a different scope for detection and monitoring. Some clients are only interested in monitoring key assets such as servers, while others value monitoring user devices such as laptops. Finally, it should be noted the collaborating SOC has a very strict definition of an `incident', i.e., an event indicating that at least one attack step against a customer asset was successful. Most  unsuccessful attacks are \textit{not} reported as incidents to customers.

For our evaluation, we obtain data collected since June 2022 on two organizations monitored by the SOC. These clients differ in terms of organization size, sector, and maturity level of their security management. One organization is a large educational institution with over $50k$ internal IP addresses responsible for over $28M$ alerts. The second organization is an medium-sized enterprise IT service company with slightly less than $5k$ internal IP addresses responsible for almost $500k$ alerts. For the first organization, the sensor setup primarily monitors crucial assets, including DNS servers, causing over $90\%$ of the traffic to be DNS connections.

\subsubsection{Alert data}
\looseness=-1The SOC uses common tools such as Zeek and Suricata to automatically generate log and alert data. The ruleset used by the SOC is gathered from various public and commercial sources in addition to their in-house developed signatures. An example of a public ruleset is Proofpoint's ET OPEN ruleset \cite{ET-Open}.
The SOC ruleset is updated daily to ensure that detection capabilities are in place to deal with emerging threats.
Moreover, following a process referred to as \textit{tuning}~\cite{alert-alchemy}, certain rules are suppressed from the ruleset because they were considered to produce too many FPs when weighted against the increased detection capabilities they may offer. To maintain external validity, we include rules originally suppressed in the SOC in our analysis.
The SOC has developed a mapping of Suricata alerts to MITRE ATT\&CK~\cite{mitre-taxonomy} techniques to improve the effectiveness of analysts during their investigation.
$\sim 26\%$ of the inspected rules cannot be unambiguously mapped to a technique.

\subsubsection{Rule dataset derivation from alert and incident data}
Based on the daily alert data, we extracted statistics that describe an organization on each day. These aggregate statistics do not contain any personal information about users. In addition, from the raw alert data, we extract a dataset containing all rule revisions that triggered an alert at the SOC and merge them with a snapshot of the ruleset. The collection of different revisions of the triggered rules allows us to evaluate the specificity of rule variations. If the investigation of an alert led to an incident report, the alert was deemed sufficiently relevant to be escalated to the client organization, and therefore can be considered a True Positive according to the SOC. To determine how many incidents were detected by each rule, we (through the collaborating SOC) automatically parse incident reports (which follow a template) using regular expressions to extract incident identifiers such as the date of the incident and the involved hosts. To check how many incidents were detected by a rule, the SOC then correlates raised alerts with incident identifiers.

\subsection{Data Description}
\begin{figure}[bp]
    \vspace{-0.15in}
    \includegraphics[width=0.8\linewidth]{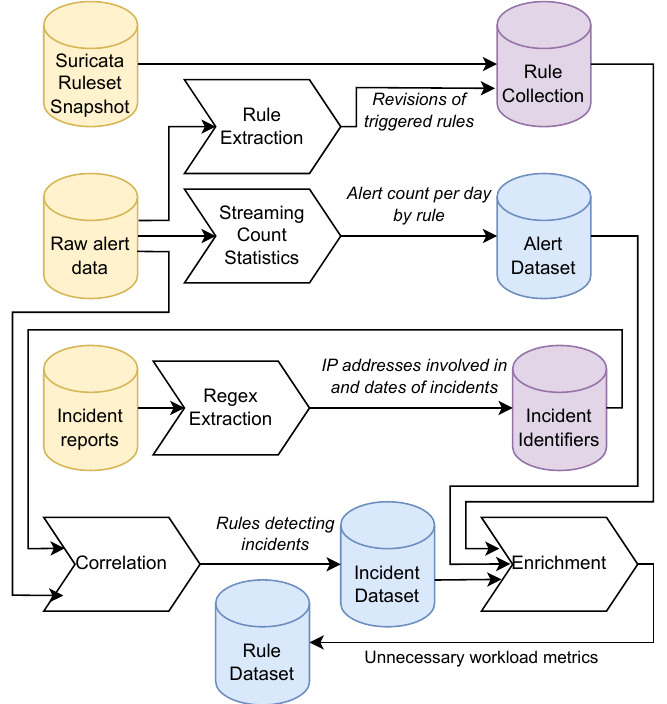}
    \caption{An information flow diagram depicting the processing steps performed to derive the three datasets describing rules, alerts, and incidents from the raw data.}
    \label{fig:data_flow}
    \Description{An information flow diagram depicting the processing steps performed to derive the three datasets describing rules, alerts, and incidents from the raw data.}
\vspace{-0.10in}
\end{figure}

\looseness=-1To evaluate rule specificity and coverage, we need a rich dataset describing how much noise rules cause and how many incidents they detect, which we obtain by combining data from several sources.
After aggregation and anonymization of the data, we obtain three datasets whose shared key statistics are summarized below and in Table~\ref{tab:data_overview} contained in Appendix~\ref{sec:appendix_1.25}. Together, the datasets describe $\sim 30M$ alerts over a period of $390$ days since June 2022 for which data is available. Over this period, we do not observe any significant change in the overall distribution of the data. The first dataset, derived from snapshots of the ruleset and the alert data, contains $\sim 290k$ rule revisions used by the SOC over time. We consider rule revisions in isolation, even if some triggering rules had more than one revision, as we find that most of the observed rule updates correspond to metadata updates, in line with \cite{ruling-the-rules}. This dataset will be enriched with using the other datasets as described in Section~\ref{sec:rule_metrics}, resulting in rich data for triggered rules. The second dataset, derived from the alert data, describes which rules triggered how often on which day. The third dataset, derived from incident data covering $42$ incidents during the same period, describes when incidents occurred and which rules contributed to the detection of those incidents. The SOC has indicated, based on the interactions with their customers, that they did not fail to report any security incident within their monitoring scope, suggesting there are no False Negatives in the incident data. The data provisioning is summarized in Figure~\ref{fig:data_flow}.

\subsubsection{Rule metrics}
\label{sec:rule_metrics}
\looseness=-1As rule specificity is considered the key to rule adoption~\cite{alert-alchemy}, we focus on the \textit{amount of unnecessary workload per day}, which corresponds to the average number of FPs triggered per day, as a metric for rule quality. This metric is computed separately for each revision of a rule. First, we determine during which period a rule has been active. An introduction date is determined by inspecting \textit{metadata} fields such as \textit{updated\_at}. Similarly, we look at the introduction dates of newer revisions of the same rule to determine a termination date for the revised rules. The introduction and termination dates are capped at the boundaries of the data collection period. To compute the amount of unnecessary workload per day, we count the number of alerts raised by a rule revision, subtract the number of detected incidents, and then divide it by the number of days for which a revision has been active. The resulting metric is also sensitive to rule design choices, including whether and how aggregation or throttling is performed. We refrain from defining a metric describing the number of missed attacks since no rule is expected to detect every incident, and coverage is provided by the ruleset as a whole.

\subsection{Design Principles Derivation}
\label{sec:methodology_design-principles}
\begin{figure}[bp]
\vspace{-0.30in}
    \includegraphics[width=0.8\linewidth]{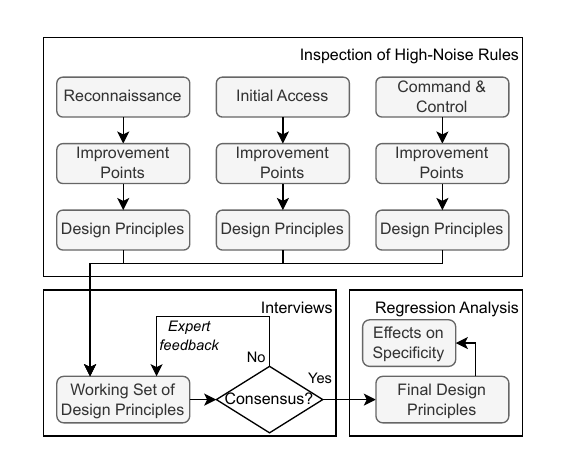}
\vspace{-0.15in}
    \caption{An overview of how the different parts of the methodology relate.}
    \label{fig:methodology_flow}
    \Description{An overview of how the different parts of the methodology relate.}
\vspace{-0.10in}
\end{figure}

The design principles presented in Section~\ref{sec:design-principles} are derived through a systematic examination of rules associated with high workload, hence lacking specificity, and validated through a series of one-on-one interviews with two domain experts. An overview of how the different parts of the methodology are related is presented in Figure~\ref{fig:methodology_flow}, and the individual steps are explained hereafter.
\subsubsection{Examination of highest workload rules}
To derive the initial set of design principles, we follow a process akin to `thematic saturation' whereby manual inspection and coding of items of interest continues until no new `codes' or `themes' emerge~\cite{saunders2018saturation}.
We select three common techniques from the three most common tactics (based on the triggered rules) to obtain a set of design principles that is generally applicable. Specifically, we focus on MITRE Techniques~\cite{mitre-taxonomy} T1595 (Active Scanning), T1190 (Exploit Public-Facing Application), and T1071 (Application Layer Protocol). 
Following the thematic saturation approach, the main researcher inspected high-noise rules and derived several `improvement points' that, in his opinion, could have increased the specificity of the rule given the threat it detected. Those improvement points are then discussed with the other authors, leveraging 15 years of intrusion detection experience in industry and academia. The process was stopped when $10$ consecutive rule inspections for that technique group did not lead to any new `improvement points' to add to the set (i.e., the set reached saturation).
The overall process required the inspection of $54$ rules. The obtained `improvement points' are then categorized and grouped to derive generally applicable rule design principles.

\subsubsection{Interviews}
\label{sec:methodology:interviews}
These proposed rule design principles were initially validated through a series of one-on-one interviews with two domain experts; one is a security researcher and one is a senior Tier-2 analyst at the SOC, with, respectively, 12 and 4 years of experience with rule engineering. The interviews were conducted independently with each expert. The focus of the interviews is to ensure the completeness and correctness of the derived design principles. During each conducted interview, we iterate over the posed design principles to check for agreement with the principle, and then ask the expert whether any other consideration should be added. The interviews were repeated until every proposed consideration was reviewed and agreed upon by both domain experts. In total, only three interviews were necessary to reach full agreement.

\subsection{Effect Size Evaluation}
\label{sec:methodology:effect-size-evaluation}
\looseness=-1We quantitatively assess the effect on unnecessary workload as described in Section~\ref{sec:rule_metrics} of (non-)conformity with design principles in a rule.
To do so, we label a selection of rules and perform a set of statistical tests to estimate effect size.
For regression, we identify groups of rules that are similar in terms of the type of threat they aim to detect but differ in their design.
This allows us to contrast workload generated by rules with and without specific design principles in place, while controlling confounding factors related to, for example, how common specific network events or traffic over a specific protocol are.
Since we aim at identifying factors influencing unnecessary workload generated by rules, high-noise rules are not excluded from the comparison groups unless the noise generated by a rule is considered to be specific to an environment and cannot be expected to be present in most similar network environments.
In forming groups, we are limited by rules that cover threats that could be observable given the sensor setup, the underlying monitored network, and the corresponding threat landscape.
We find that over $70\%$ of all rules in the ruleset and over $50\%$ of the triggered rules operate on the \textit{http} protocol.
Therefore, all groups are constrained to rules that perform detection on top of the \textit{http} protocol, which is representative of the majority of rules.
To ensure homogeneous groups, we select them based on MITRE Techniques~\cite{mitre-taxonomy} and only consider groups based on techniques with more than $15$ rules.
We label each rule with the implemented design principles.

\looseness=-1One group of rules consists of $21$ reconnaissance rules detecting technique T1595 (Active Scanning). Note that the rule discussed in Section~\ref{sec:worst-offenders:c2}, which may detect outbound scans initiated by an infected host, was excluded from this group due to its high environment-specific noise.
Another group consists of $9$ Initial Access rules detecting behavior in relation to Log4j~\cite{apache-log4j-cve} exploitation, comprising rules including the string \textit{Log4j} in the alert message. The Log4j group is especially interesting since the SOC uses rules from multiple distinct sources with different designs to maximize coverage due to the associated high risk, which enables a comparison of the effects of different designs.
We complement the Log4j group with another group covering the same technique but using different procedures, which consists of $81$ rules detecting technique T1190 (Exploit Public-Facing Application), and excludes all rules from the Log4j group.
Further Initial Access groups consist of $53$ rules detecting technique T1055 (Process Injection) and $19$ rules detecting T1189 (Drive-by-Compromise).
The last group consists of $19$ Command \& Control rules, which are selected based on their technique, T1071 (Application Layer Protocol).
Together, the groups include $182$ rules.

To evaluate the effect size, we rely on a batch of robust linear regressions using statsmodels~\cite{statsmodels-paper, statsmodels}, estimating the effect of any given design principle on the expected daily workload. We do not regress over design principles for which a group has less than two rules (not) adhering to that design principle. Following common practice~\cite{variance-inflation-factor}, we assert that the Variance Inflation Factors (VIF) do not exceed $20$ for all coefficients before performing multiple regression. We report statistical significance for each estimated effect size from a batch of t-tests; p-values lower than $0.05$ are considered significant. In addition, we use the Kolmogorov-Smirnov Test (KS test) as a goodness-of-fit test to better understand how much variance can be explained by the model~\cite{kolmogorov-smirnov-kstest}. Rejection of the null hypothesis ($H_0$) of a KS-test implies poor fit. The KS test is chosen over the usual $R^2$ due to its robustness against outliers.

\subsection{Ethical Considerations}
\label{sec:ethics}
This research was carried out with ethical approval from our institution’s ethical review board under approval number \anon{ERB2024MCS01}. We gained explicit and informed consent from all subjects participating in the interviews, and subjects were assured the study would not affect their working conditions in any way. Sensitive log, alert, and incident data have, where relevant, been aggregated and/or anonymized at the SOC, such that researchers cannot use those data to identify systems or individual people to whom the data relates.
The interviewees explicitly consented to their participation in this research after receiving sufficient information on the topic and remain anonymous. Concerning the evasion of detection, we have taken care not to discuss potential evasion strategies relating to vulnerabilities that are recent and still likely to be exploitable.

\section{Data Exploration}
\label{sec:data_exploration}
\looseness=-1Section~\ref{sec:data_exploration:typical_rules} describes at a high level the ruleset and the typical rules contained. Subsequently, Section~\ref{sec:data_exploration:incident_detection} provides an overview of the relation between rules, alerts, and incidents, highlighting which types of rules contribute to detecting incidents. Lastly, Section~\ref{sec:worst-offenders} presents three high-noise rules to exemplify the design principle derivation.

\subsection{Ruleset Overview \& Typical rules}
\label{sec:data_exploration:typical_rules}
\begin{figure}[bp]
    \includegraphics[width=1.0\linewidth]{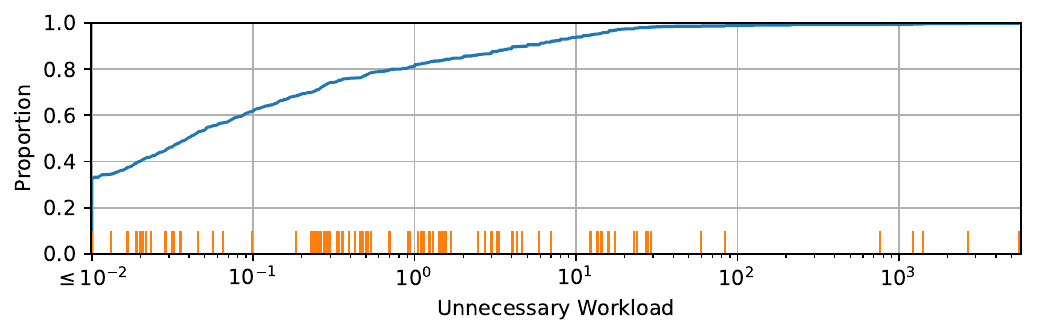}
\vspace{-0.30in}
    \caption{An Empirical Cumulative Distribution Function Plot for the unnecessary workload (\textit{x-axis, log-scale}) generated by triggered non-informational rules. The markers on the \textit{x-axis} represent rules contributing to detecting incidents.}
    \label{fig:unnecessary_workload_incidents_ecdf}
    \Description{An Empirical Cumulative Distribution Function Plot for the unnecessary workload (\textit{x-axis, log-scale}) generated by triggered non-informational rules. The markers on the \textit{x-axis} represent rules contributing to detecting incidents.}
\vspace{-0.10in}
\end{figure}

Based on our initial interactions with the SOC, we gained a better understanding of what SOC managers and analysts demand from intrusion detection rules and the context in which these operate. According to a SOC representative, some rules have such a high false positive rate that they are considered to be \textit{informational} and only used for threat hunting or as supporting evidence for an investigation initiated based on another alert. The ruleset used consists of approximately $290k$ unique rules, of which approximately $9k$ are considered by the SOC to be informational. Since non-informational alerts are responsible for the majority of the analyst workload and detection coverage, we focus on non-informational rules in the remainder of this work.
Among these, $791$ rules were triggered during the period for which we have data. Hence, a significant portion of the ruleset never triggers. Figure~\ref{fig:unnecessary_workload_incidents_ecdf} shows the distribution of unnecessary workload generated by triggered non-informational rules and highlights the workload generated by rules that contribute to the detection of incidents using markers on the \textit{x-axis}. The figure suggests that a minority of rules causes significant noise while low-noise rules can effectively contribute to detecting incidents.

\subsection{Incident Detection by Rules}
\label{sec:data_exploration:incident_detection}
Although the SOC has only reported $42$ incidents during the $390$ days for which we have full data, they had accumulated several millions of alerts over the same period. We find that the majority of alerts are not interesting for SOC analysts to investigate. During our analysis of Figure~\ref{fig:unnecessary_workload_incidents_ecdf}, we found that approximately $80\%$ of the triggered rules raise less than a single alert per day. However, a very small fraction of the rules is responsible for the majority of unnecessary workload that causes alert fatigue.

\begin{table}[tp]
    \caption{A breakdown of rules contributing to the detection of incidents by rule category.}
\vspace{-0.15in}
    \label{tab:detected_incidents_by_rule_category}
    {\small
    \begin{tabular}{lr}
        \toprule
        Total No. Detected Incidents & 42 \\
        \midrule
        Rule Category & No. Detected Incidents \\
        \midrule
        Content & 17 \\
        Threshold & 7 \\
        DNS & 32 \\
        IP & 4 \\
        Other & 3 \\
        \bottomrule
    \end{tabular}
    }
    \Description{A breakdown of rules contributing to the detection of incidents by rule category.}
\vspace{-0.25in}
\end{table}
Table~\ref{tab:detected_incidents_by_rule_category} shows a breakdown of the categories of rules that contribute to the detection of incidents. Note that some incidents are detected by multiple categories of rules, causing the sum of incidents over the different categories to be larger than the total number of incidents. Over $75\%$ of incidents are detected through the usage of rules that detect specific domain names as Indicators of Compromise (IoC) in a DNS request. The low detection rate for rules leveraging the content of packets opposite to atomic IoCs may be attributed to the sensor setup of the largest customer of the SOC, where DNS servers are monitored, but not inbound/outbound traffic of less critical hosts within the monitoring scope. Therefore, we refrain from suggesting content-based detection strategies beyond atomic IoC are ineffective in detecting incidents.

\looseness=-1Next, we turn to the relation between rule age and coverage. $7$ incidents out of $42$ incidents for which we have data were detected by recently introduced rules (1-7 days old). For $4$ of these incidents, such a recently introduced rule is the only rule that contributes to the detection of that incident. This suggests that it is important for SOC managers to ensure regular updates of the employed ruleset. Furthermore, $10$ incidents were only detected by legacy rules, which were last updated more than $2$ years before the incident and of which the oldest was not updated for over $9$ years. Although some of these rules correspond to generic threats such as network scans, others correspond to C2 traffic from older malware such as Zeus~\cite{talos-zeus-dga} (see Figure~\ref{fig:zeus-dga}).

\looseness=-1This rule is a good example of a specific rule that still achieves general coverage. Even if this rule was last updated in $2014$, it barely caused any workload and has contributed to the detection of incidents. This legacy rule detects domains corresponding to the Zeus botnet DGA algorithm~\cite{talos-zeus-dga} and potentially other types of malware using DGA strategies. The exact DGA domain has been generalized to a structure and is matched at variable offset using the \textit{pcre} feature from Suricata. Leveraging the incident reports, we establish the rule did not detect a Zeus-related infection but did detect another type of infection employing DGA techniques similar to Zeus's. To reduce FPs, alert throttling is applied in a way that alerts are only raised after several observations, and only one alert is generated every two minutes. Repetitive queries to DGA domains would indicate an internal host has already been successfully infected with malware.
\begin{figure}[htb]
\begin{lstlisting}
alert udp any 53 -> $HOME_NET any (
msg:"ET MALWARE Possible Zeus GameOver/FluBot Related DGA NXDOMAIN Responses";
byte_test:1,&,128,2; byte_test:1,&,1,3; byte_test:1,&,2,3;
content:"|00 01 00 00 00 01|"; offset:4; depth:6;
pcre:"/^..[\x0d-\x20][a-z]{13,32}(?:\x03(?:biz|com|net|org)|\x04info|\x02ru)\x00\x00\x01\x00\x01/Rs";
threshold:type both, track by_dst, count 12, seconds 120;)
\end{lstlisting}
\vspace{-0.15in}
    \caption{Shortened rule detecting Zeus DGA NXDOMAIN responses (sid: $2018316$)}
    \label{fig:zeus-dga}
    \Description{Shortened rule detecting Zeus DGA NXDOMAIN responses (sid: $2018316$)}
\vspace{-0.15in}
\end{figure}

\subsection{Inspection of High-Noise Rules}
\label{sec:worst-offenders}
\looseness=-1To better understand what distinguishes good rules from high-noise rules, we investigate a variety of rules known to cause mass FPs in the monitored environments following the methodology described in Section~\ref{sec:methodology_design-principles}. Looking at ($54$ of) these high-noise rules and using our knowledge of intrusion detection and Suricata~\cite{Suricata} in particular, we reason about which alternate design choices could have improved these rules. To exemplify this process, we cover three rules and areas in which they can be improved. These rules were chosen as an example because they allow us to highlight all design principles derived from this process and to cover all three different MITRE Tactics~\cite{mitre-taxonomy} we investigated. For brevity, we remove fields not affecting specificity or coverage, such as \textit{metadata} and \textit{fast\_pattern}.

\subsubsection{Reconnaissance}
\label{sec:worst-offenders:recon}
\begin{figure}[htb]
\vspace{-0.15in}
\begin{lstlisting}
alert http $EXTERNAL_NET any -> $HOME_NET any (
msg:"ET SCAN OpenVAS User-Agent Inbound";
flow:established,to_server;
http.user_agent; content:"OpenVAS";)
\end{lstlisting}
\vspace{-0.15in}
    \caption{Shortened rule detecting inbound OpenVAS UA (sid: $2012726$)}
    \label{fig:rule-openvas}
    \Description{Shortened rule detecting inbound OpenVAS UA (sid: $2012726$)}
    \vspace{-0.15in}
\end{figure}

\looseness=-1The rule shown in Figure~\ref{fig:rule-openvas} detects the OpenVAS HTTP User-Agent (UA) in inbound traffic. OpenVAS is an open-source vulnerability scanning tool~\cite{openvas} and has legitimate use cases for penetration testers who must assess the attack surface of organizations. However, these tools are also abused by malicious actors~\cite{sentinelone-openvas, reliaquest-openvas} to find vulnerable systems to exploit. Internet-connected systems are likely to be periodically scanned by infected hosts~\cite{internet-background}, and for certain network environments monitored by the SOC described in Section~\ref{sec:data_sources}, this will result in many alerts. Due to the broadness of this rule, generated alerts are not very actionable since it is unclear what vulnerability was scanned for. More pressingly, given an alert, it is unclear whether an exploitable vulnerability was detected.

We envision two potential avenues for improving this rule. A first improvement opportunity for this rule is to make it more specific with respect to the type of vulnerability scanned for and to include checks on the server response to assess whether a vulnerability was detected or not. This approach helps distinguish between scan attempts from which nothing of interest was learned and successful scans from which adversaries may proceed to exploit a system. If coverage is preferred over or desired in conjunction with specificity, it is possible to apply alert throttling to the rule, effectively limiting the number of alerts to one per scanner within a certain timeframe. By correlating an alert generated by the alert throttled rule with other network logs, such as Zeek logs, an analyst can still determine exactly which hosts and services were scanned.
\subsubsection{Initial Access}
\label{sec:worst-offenders:initial-access}
\begin{figure}[htb]
\begin{lstlisting}
alert http any any -> $HTTP_SERVERS any (
msg:"ET WEB_SERVER ColdFusion administrator access";
flow:established,to_server;
http.method; content:"GET"; nocase;
http.uri; content:"/CFIDE/administrator"; nocase;)
\end{lstlisting}
\vspace{-0.15in}
    \caption{Shortened rule detecting ColdFusion unauthorized administrator access attempts (sid: $2016184$)}
    \label{fig:rule-coldfusion}
    \Description{Shortened rule detecting ColdFusion unauthorized administrator access attempts (sid: $2016184$)}
    \vspace{-0.15in}
\end{figure}
The rule highlighted in Figure~\ref{fig:rule-coldfusion} aims at detecting unauthorized administrative access to the Adobe ColdFusion software by detecting inbound HTTP requests to one of the admin URIs. Based on our conversations with the SOC regarding this rule, we conclude that the majority of FPs caused by this alert involve systems that do not run ColdFusion software and therefore cannot be vulnerable. Moreover, we note that even if a system runs the ColdFusion software, it is unclear from alerts generated by this rule whether unauthorized access was granted or denied. The rule also lacks generalization capabilities, considering that it only detects one of the several URIs related to the vulnerability~\cite{coldfusion-advisory, coldfusion-exploitdb}.

To generalize the rule to cover other URIs related to the same group of CVEs mentioned in the security advisory~\cite{coldfusion-advisory}, one could reduce the first content match to just the \textsc{/CFIDE/} substring to reduce the need for a second rule\footnote{The second rule is given in Figure~\ref{fig:rule-coldfusion-2} in Appendix~\ref{sec:appendix_1}.}. Using a regular expression, one could then match whether one of the three specific URIs mentioned in the advisory is requested. A more pressing improvement is that there appears to be a clear way to detect whether a system is vulnerable based on the response of the server~\cite{coldfusion-exploitdb2}. An HTTP $200$ status code suggests that access was granted, and hence, the system appears to be vulnerable according to the Metasploit code. Similarly to the approach proposed in Section~\ref{sec:worst-offenders:recon}, the alerts generated by this detection method could be limited to the network connections in which not just an attempt was made, but to the connections in which unauthorized access was successfully acquired.
\subsubsection{Command \& Control}
\label{sec:worst-offenders:c2}
\begin{figure}[htb]
\vspace{-0.15in}
\begin{lstlisting}
alert http $HOME_NET any -> $EXTERNAL_NET any (
msg:"ET USER_AGENTS Go HTTP Client User-Agent";
flow:established,to_server;
http.user_agent; content:"Go-http-client"; nocase;)
\end{lstlisting}
\vspace{-0.15in}
    \caption{Shortened rule detecting outbound requests with anomalous HTTP UA (sid: $2024897$)}
    \label{fig:rule-go-user-agent}
    \Description{Shortened rule detecting outbound requests with anomalous HTTP UA (sid: $2024897$)}
\vspace{-0.15in}
\end{figure}
\looseness=-1Figure~\ref{fig:rule-go-user-agent} shows a rule with the goal of detecting certain outbound traffic, likely originating from an infected machine. Hence, such traffic could be Command and Control traffic or malicious traffic aimed at spreading malware. Some malware is known to use the default HTTP UA from the Golang standard HTTP client~\cite{paloalto-malware-go, crowdstrike-malware-go, akamai-malware-go}, and hence detecting that UA offers a viable detection opportunity for a broad range of malware. However, there are benign applications that may use the same UA~\cite{tailscale-go}, and one instance of such a benign application could cause an excess of benign triggers, since the rule will generate an alert for every HTTP request with that UA. Although this example may appear rare, the SOC affirms these benign applications, contribute significantly to unnecessary workload. The usage of benign applications on which this rule raises FPs results in environment-specific noise, which is distinct from the universal noise generated by the two previously examined rules in any public unfiltered network environment.

Although there is no available documentation on the origin of this rule, we suspect a trade-off was made between coverage and specificity, and this choice led to the detection of the UA as a proxy for actions with a malicious origin, although other characteristics of the corresponding malware can be identified~\cite{akamai-malware-go}. Detection based on characteristics more closely related to malicious actions themselves can increase specificity at the cost of coverage. Another way to improve this rule is to include exceptions for common benign software (such as \cite{tailscale-go}) using the same UA by including a negative match on other characteristics within a benign HTTP request such as a header specific to the benign software and, if need be, excluding traffic towards certain known-benign IP addresses from inspection. Alternatively, alert throttling can be applied to limit the number of FPs to one per client within a certain timeframe.

\section{Principles for NIDS Rule Design}
\label{sec:design-principles}
We formulate our design principles in Section~\ref{sec:design-principles:formulation} and in Section~\ref{sec:design-principles:statistical} we use regression to assess their effect on unnecessary workload.

\subsection{Rule Design Principles}
\label{sec:design-principles:formulation}
\looseness=-1We propose six rule design principles based on our analysis. The first four are aimed at increasing rule specificity, whereas the last two are intended to ensure coverage is preserved. 
For each design principle, we highlight considerations for specific tactics when applicable. 
Our principles may offer choices resulting in rules on a scale ranging from better to worse. For rules significantly diverging from the first four principles, we propose taking action to prevent unnecessary workload. 
Specifically, one may consider applying more aggressive alert throttling, reducing the monitoring scope by ignoring certain traffic directions more prone to noise, or even labeling the rule as ``informational'' such that alerts generated by the rule should not be meant to trigger investigations, but only serve as contextual information in investigating alerts from other rules.

\subsubsection{Limited Proxy}
\label{sec:design-consideration-no-proxy}
Here, we deal with the choice of the characteristic upon which detection is based. Ideally, rules should detect a characteristic that is observable as a direct consequence of the malicious behavior, such as a malicious payload. This may not be possible for all types of malicious behavior due to encryption, limitations of the detection engine, or similarity to benign behavior. In those cases, it may be possible to detect a proxy for the malicious behavior (e.g., a UA set by malware). 
A characteristic that can be used as a proxy and does not occur in benign behavior may not exist for certain types of malicious behavior. In those cases, one may resort to detecting anomalous behavior that is correlated with malicious behavior and is believed to rarely be benign.

Using characteristics relating directly to the malicious behavior will result in lower-noise rules. As an additional advantage, these rules are typically easier to interpret since they point directly to malicious behavior. The use of proxies may offer more opportunities for detection, but may also result in increased noise due to benign triggers, as we have seen in Section~\ref{sec:worst-offenders:c2}. 
The uniqueness of proxies to malicious behavior plays a determining role. It should be noted that estimating how rare certain benign behaviors are is notoriously difficult. Rules detecting anomalous behavior may therefore result in a significant amount of noise in certain environments.

Due to the increased adoption of encryption in network traffic (also by malware), it has become increasingly difficult to find suitable characteristics for detection~\cite{encryption-blessing-or-curse}. A common group of proxies leveraged for detection are JA3 and JA3s hashes~\cite{ja3}, which are hashes of server/client TLS configurations. During our interviews, it became evident that JA3(s) hashes are useful but have limitations. It can be difficult to establish how common certain JA3(s) hashes are in benign traffic. Although it may be possible to scan common internet services to discover common JA3s hashes~\cite{ja3-server-reliability}, it can be much more challenging to do something similar for client hashes, although attempts have been made~\cite{ja3-list-client}.

\noindent\emph{Application Example.} 
Considering the Reconnaissance rule for OpenVAS discussed in Section~\ref{sec:worst-offenders:recon}, we could improve it by replacing the OpenVAS UA as a characteristic by a Base64-encoded test string included in the request by OpenVAS for a specific vulnerability scan.
A rule example is provided in Appendix~\ref{sec:appendix_1}, Figure~\ref{fig:rule-openvas-better}.

\subsubsection{Successful Malicious Action}
\label{sec:design-consideration-successful}
Ideally, rules should trigger only if a malicious action was successfully executed.
To achieve this, sometimes it may be necessary to observe an additional traffic portion. 
From this second portion, one may conclude that a malicious action was successful or that a malicious action could not have been successful. 
The second traffic portion may typically be observed in a server response or a follow-up connection.

However, there are exceptions.
One may still accept to build a rule potentially detecting unsuccessful malicious behavior if that behavior poses a high risk.
We can distinguish between two types of high-risk behavior, namely temporal and structural risk. The temporal risk may, for example, be high directly after discovering a zero-day vulnerability. The structural risk may be high if vulnerabilities are easy to exploit and have a high impact. Contrary to structural risk, temporal risk changes over time and, as a result, rules may require updates when temporal risk changes.

Additionally, an unsuccessful malicious action may still indicate that a previous malicious action has already succeeded, as would be the case for lateral movement attempts. Hence, the monitoring scope in which the malicious behavior is observed is important. 
A network scan or initial access attempt executed by an internal host is more interesting to be alerted about than a scan executed by an arbitrary host, since it may indicate the internal host is compromised. Alternatively, it should be possible to establish whether a scan was consequential by leveraging observations in Section~\ref{sec:worst-offenders:recon}. 

During the design of a rule, one should also be mindful of the sensor setup and the impact that the setup has on the monitoring scope. For instance, if an internal host makes a DNS query for a website hosting an exploit kit domain, such requests are more likely to correspond to a successful infection of that host if the host successfully follows up on the DNS query by making a request and receiving a response from the host to which the domain name resolved. 
It is a common strategy to center detection efforts around an organization's crown jewels and only monitor traffic to and from that limited group of hosts~\cite{mitre-world-class-soc}. As a result, it may only be possible for a SOC to detect a malicious DNS request but not the corresponding follow-up connection. As a rule of thumb, on the safe side, one may assume that only follow-up connections between the same pair of hosts can be guaranteed to be observable.

\noindent\emph{Application Example.} 
\looseness=-1The Initial Access rule discussed in Section~\ref{sec:worst-offenders:initial-access} can be improved using {\small\texttt{flowbits}} to check for an HTTP 200 response. Concretely, {\small\texttt{flowbits: set, coldfusion\_admin\_access; flowbits: noalert;}} is added to the rule shown in Figure~\ref{fig:rule-coldfusion} and a new rule is written to check the response, which is shown in Figure~\ref{fig:rule-coldfusion-addition}.

\begin{figure}[htb]
\vspace{-0.15in}
\begin{lstlisting}
alert http $HTTP_SERVERS any -> any any (
msg:"ET WEB_SERVER ColdFusion successful administrator access";
flow:established,to_client;
flowbits: isset, coldfusion_admin_access;
http.stat_code; content:"200";)
\end{lstlisting}
\vspace{-0.15in}
    \caption{Shortened rule detecting successful Coldfusion administrator unauthorized access jointly with the modified version of the rule in Figure~\ref{fig:rule-coldfusion} as described in Section~\ref{sec:design-consideration-successful}}
    \label{fig:rule-coldfusion-addition}
    \Description{Shortened rule detecting successful Coldfusion administrator unauthorized access jointly with the modified version of the rule in Figure~\ref{fig:rule-coldfusion} as described in Section~\ref{sec:design-consideration-successful}}
    \vspace{-0.15in}
\end{figure}

\subsubsection{Alert Throttling}
\label{sec:design-consideration-threshold}
Alerting should be limited in a way permitting the identification of attacking devices and (potentially) compromised devices.
Moreover, different steps of an incident should be identifiable, though this can usually be achieved through a combination of different rules covering different steps. 
If alert throttling cannot be applied such that the distinction of different steps is possible, it might be necessary to split a rule into several more specific ones, especially if different steps may have different implications.

\looseness=-1Usually, rules can be throttled to generate at most one alert for each combination of source and destination IPs in a given time window. The duration of the time window can be chosen to differentiate repeated actions to the extent they may be relevant.
Since we consider building rules in a context where alert data is supported by generic traffic logs (see Section~\ref{sec:background_soc}), it is still possible to derive how often and when a certain traffic pattern was observed, given one example of this pattern. 
For rules detecting reconnaissance, such as the rule analyzed in Section~\ref{sec:worst-offenders:recon}, it is possible to apply an aggressive throttling strategy so that at most one alert is generated per source.

\noindent\emph{Application Example.} Considering the reconnaissance rule presented in Section~\ref{sec:worst-offenders:recon},
a lower noise alternative based on alert throttling is shown in Appendix~\ref{sec:appendix_1}, Figure~\ref{fig:rule-openvas-better} and adds {\small\texttt{threshold: type limit, track by\_src, count 1, seconds, 60;}}. 
This improved version will trigger, at most, one alert per source IP address every minute.

\subsubsection{Exceptions}
\label{sec:design-consideration-exceptions}
Rule developers should devote effort to identifying in which circumstances a rule may trigger on benign traffic and creating exceptions for it.
This is especially relevant for rules that trigger on a proxy for malicious behavior or on anomalous behavior (see Section~\ref{sec:design-consideration-no-proxy}). 
The interviewees indicated that it is common that some rules work perfectly in some environments but can cause significant unnecessary workloads in others, as we could conclude from our observations in Section~\ref{sec:worst-offenders:c2}. This is also confirmed by exceptions covering common antivirus software in some rules we inspected~\footnote{An example of such rule is provided in Figure~\ref{fig:executable-download-exceptions} in Appendix~\ref{sec:appendix_1}.}. Such software may be absent in many organizations and omnipresent in others, resulting in many benign triggers.
Although some exceptions may be generic, others may be environment-specific similar to the noise they deflect. An additional consideration is that if an adversary knows the deployed rules and their exceptions, it may allow the attacker to evade detection if the exception condition is a characteristic they can control.

\noindent\emph{Application Example.} 
Considering the Command \& Control rule discussed in Section~\ref{sec:worst-offenders:c2}, for which we discussed a common FP. The rule could be improved by adding a negative content match on a request header that is specific to the benign software that caused FPs~\footnote{\url{https://github.com/tailscale/tailscale/blob/f1d10c12acf69fcb7509c6aaff65e1a9c8897715/net/netcheck/netcheck.go\#L1034}}. 
The Suricata implementation for such a match would be:\\
{\small\texttt{http.request\_header; content:!"X-Tailscale-Challenge|3a 20|";}}.

\subsubsection{Generalized Characteristic}
\label{sec:design-consideration-generalized}
As discussed in Section~\ref{sec:worst-offenders:initial-access}, characteristics for detection can be chosen in a way to provide coverage for different variations of similar threats. From malware analysis reports, including one report we will use as an example covering DarkVNC~\cite{reaqta-darkvnc}, we know that characteristics that can be leveraged for detection may vary depending on the infected host and even depending on randomness.
Specifically, one may consider each characteristic to have invariable and variable aspects. The hostname and the randomized identifier observable as part C2 traffic can be considered to be \textit{variable}, although the format in which these are observed or specific parts (e.g., the \textit{DarkVNC} string) can both be considered to be \textit{invariable} since these are more likely to be consistent across different incidents. Rule developers should prefer matching invariable parts to promote coverage.

\noindent\emph{Application Example.} The initial access rule discussed in Section~\ref{sec:worst-offenders:initial-access} could be improved by replacing the content match \\ {\small\texttt{http.uri; content:"/CFIDE/administrator"; nocase;}} 
by a shorter one and a regular expression using \textit{pcre}. 
Concretely, the content match would be replaced by:
{\small\texttt{http.uri; content:"/CFIDE/"; pcre: "/\textbackslash/CFIDE\textbackslash/(administrator|adminapi)/i";}}.

\subsubsection{Generalized Position}
\label{sec:generalized-position}
In addition to the characteristic being matched by a rule, one should also consider how and specifically at which location in the network traffic these characteristics are matched. Adversaries may execute variations of known attacks, potentially with the aim of evading detection, causing the same characteristic to be observed at a different location. 
As an example, consider a vulnerable web application and a rule~\footnote{An example of such rule is provided in Appendix~\ref{sec:appendix_1}, Figure~\ref{fig:command-injection-generalized-location}.} using \texttt{content:"/login.cgi?cli=";} as detection option. 
This rule would not trigger if instead an attacker would have crafted a request such as \texttt{/login.cgi?foo=bar\&cli=}. 
Splitting the detection logic into two parts to separately detect the path and the name of the query parameter is a way to generalize coverage of the rule whilst still leveraging the same characteristic.

\noindent\emph{Application Example.} The aforementioned detection option would be replaced by \texttt{content:"/login.cgi?"; content:"cli=";}.

\subsection{Effects of Design Principles on Specificity}
\label{sec:design-principles:statistical}

\paragraph{Comparison Groups}
\looseness=-1As mentioned in Section~\ref{sec:methodology_design-principles}, we consider six groups, consisting of $182$ rules in total, that are homogeneous in the sense that the rules contained therein detect similar behavior and are diverse in the sense that they are designed in a different way so that we can compare the effect of design choices made.
We manually assign a binary label to each rule concerning the design principles described in Section~\ref{sec:design-principles:formulation}. For certain design criteria and groups, all rules were assigned the same label for a design principle because no rules accounted particularly well/poorly for that principle, causing no regression to be performed over these variables. Table~\ref{tab:labels_distribution} shows an overview of how labels are distributed. We found that rules detecting Initial Access techniques rarely distinguish between successful and unsuccessful actions and rarely use alert throttling.
\begin{table*}[tp]
    \caption{Breakdown of the different groups and the decomposition of their labels for the prevalence of various design principles.}
\vspace{-0.15in}
    \label{tab:labels_distribution}
    {\small
    \begin{tabular}{lrrrrrr}
        \toprule
        & \multicolumn{6}{c}{Design Principle} \\
        Comparison Group & Limited Proxy & \shortstack{Successful \\ Malicious Action} & Exceptions & Alert Throttling & Generalized Characteristic & Generalized Position\\
        \midrule
        Active Scanning     & $95\%$ ($20$)     & $5\%$ ($1$)   & $5\%$ ($1$)   & $10\%$ ($2$)  & $38\%$ ($8$)      & $48\%$ ($10$) \\
        Log4J               & $100\%$ ($9$)     & $22\%$ ($2$)  & $0\%$ ($0$)   & $67\%$ ($6$)  & $22\%$ ($2$)      & $100\%$ ($9$) \\
        Other Exploits      & $93\%$ ($75$)     & $0\%$ ($0$)   & $2\%$ ($2$)   & $0\%$ ($0$)   & $33\%$ ($27$)     & $89\%$ ($72$) \\
        Process Injection   & $100\%$ ($33$)    & $0\%$ ($0$)   & $3\%$ ($1$)   & $0\%$ ($0$)   & $15\%$ ($5$)      & $97\%$ ($32$) \\
        Drive-by-Compromise & $74\%$ ($14$)     & $11\%$ ($2$)  & $5\%$ ($1$)   & $0\%$ ($0$)   & $26\%$ ($5$)     & $89\%$ ($17$) \\
        Command \& Control  & $5\%$ ($1$)       & $79\%$ ($15$) & $16\%$ ($3$)  & $5\%$ ($1$)   & $79\%$ ($15$)     & $26\%$ ($5$) \\
        \bottomrule
    \end{tabular}
    }
    \Description{Breakdown of the different groups and the decomposition of their labels for the prevalence of various design principles.}
\vspace{-0.05in}
\end{table*}
\paragraph{Results}
\label{sec:design-principles:statistical:results}
\begin{table*}[tp]
    \caption{Effects of the design principles on the volume of unnecessary workload generated per day and their significance. Significant coefficients and corresponding p-values are typeset in \textit{bold}. \textit{Dashes (-)} are used to indicate no regression was performed over a label due to a lack of positively/negatively labeled samples for that label within that group. At the bottom of the table, the number of observations is shown, as well as the test statistic and p-values for the Kolmogorov-Smirnov tests conducted as goodness-of-fit tests. The figure is a visual representation of the table showing the regression coefficients together with the corresponding Confidence Intervals (CIs).}
    \vspace{-0.15in}
    \label{tab:regression_workload}
    \label{fig:regression_visualization}
\begin{minipage}{\textwidth}
{\small
\begin{tabular}{lrrrrrrrrrrrr}
\toprule
& \multicolumn{12}{c}{Design Principle} \\
Comparison Group & \multicolumn{2}{c}{\shortstack{Active \\ Scanning}} & \multicolumn{2}{c}{Log4j} & \multicolumn{2}{c}{\shortstack{Other \\ Exploits}} & \multicolumn{2}{c}{\shortstack{Process \\ Injection}} & \multicolumn{2}{c}{\shortstack{Drive-by\\-Compromise}} & \multicolumn{2}{c}{\shortstack{Command \\ \& Control}} \\
& Coef. & P-val. & Coef. & P-val. & Coef. & P-val. & Coef. & P-val. \\
\midrule
Constant                                                & $0.07$          & $0.87$          & $\mathbf{7.45}$                               & $\mathbf{<0.01}$ & $\mathbf{0.51}$\textsuperscript{\circled{1}}  & $\mathbf{<0.01}$ & $\mathbf{0.04}$          & $\mathbf{<0.01}$           & $\mathbf{0.06}$                               & $\mathbf{<0.01}$ & $0.18$                                        & $0.10$           \\
Limited Proxy                                           & $-$             & $-$             & $-$                                           & $-$              & $\mathbf{-0.41}$\textsuperscript{\circled{2}} & $\mathbf{<0.01}$ & $-$             & $-$              & $-${\footnotesize \textsuperscript{$\ast$}} & $-$ & $-$                                           & $-$              \\
Successful Malicious Action                             & $-$             & $-$             & $\mathbf{-2.71}$\textsuperscript{\circled{3}} & $\mathbf{<0.01}$ & $-$                                           & $-$              & $-$             & $-$              & $\mathbf{-0.04}$\textsuperscript{\circled{4}} & $\mathbf{<0.01}$ & $\mathbf{-0.19}$\textsuperscript{\circled{5}} & $\mathbf{0.02}$  \\
Exceptions\textsuperscript{\circled{9}}                & $-$             & $-$             & $-$                                           & $-$              & $-0.04$                                       & $0.64$           & $-$         & $-$           & $-$                                           & $-$              & $-0.02$                                       & $0.82$           \\
Alert Throttling                                        & $0.58$          & $0.50$          & $\mathbf{-7.39}$\textsuperscript{\circled{6}} & $\mathbf{<0.01}$ & $-$                                           & $-$              & $-$             & $-$              & $-$                                           & $-$              & $-$                                           & $-$              \\
Generalized Characteristic\textsuperscript{\circled{7}} & $\mathbf{1.27}$ & $\mathbf{0.02}$ & $\mathbf{0.82}$                               & $\mathbf{<0.01}$ & $-0.02$                                       & $0.61$           & $0.02$          & $0.44$           & $\mathbf{1.87}${\footnotesize \textsuperscript{$\ast$}}                               & $\mathbf{<0.01}$ & $0.05$                                        & $0.51$           \\
Generalized Position\textsuperscript{\circled{8}}       & $0.43$          & $0.38$          & $-$                                           & $-$              & $-0.00$                                       & $0.96$           & $-$         & $-$           & $-0.01$                                        & $0.26$           & $0.04$                                        & $0.61$           \\
\midrule
N.obs. & \multicolumn{2}{r}{$21$} & \multicolumn{2}{r}{$9$} & \multicolumn{2}{r}{$81$} & \multicolumn{2}{r}{$33$} & \multicolumn{2}{r}{$19$} & \multicolumn{2}{r}{$19$} \\
Kolmogorov-Smirnov Test                                 & $0.33$          & $0.20$          & $0.22$                                        & $0.99$           & $\mathbf{0.58}$                               & $\mathbf{<0.01}$ & $\mathbf{0.58}$ & $\mathbf{<0.01}$ & $0.32$                                        & $0.31$           & $\mathbf{0.53}$                               & $\mathbf{<0.01}$ \\
\bottomrule
\end{tabular}
\\
\begin{flushleft}
{\footnotesize \quad \textsuperscript{$\ast$} \, The Limited Proxy principle was omitted from the regression since the Limited Proxy and Generalized Characteristic labels are the invese of eachother for the Drive-by-Compromise group.}
\end{flushleft}
}
\vspace{-0.15in}
\includegraphics[width=\linewidth]{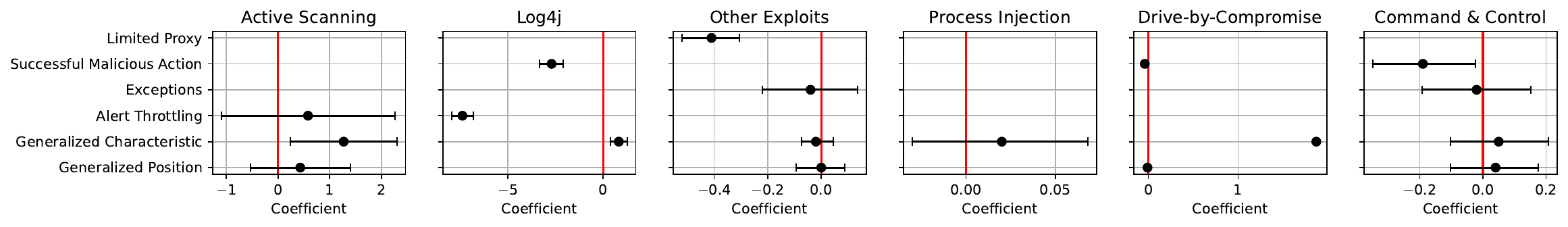}
\vspace{-0.2in}
\end{minipage}
    \Description{Effects of the design principles on the volume of unnecessary workload generated per day and their significance. Significant coefficients and corresponding p-values are typeset in \textit{bold}. \textit{Dashes (-)} are used to indicate no regression was performed over a label due to a lack of positively/negatively labeled samples for that label within that group. At the bottom of the table, the number of observations is shown, as well as the test statistic and p-values for the Kolmogorov-Smirnov tests conducted as goodness-of-fit tests. The figure is a visual representation of the table showing the regression coefficients together with the corresponding Confidence Intervals (CIs).}
    \vspace{-0.15in}
\end{table*}
Table~\ref{tab:regression_workload} shows the outcome of the regression performed over the various groups according to the methodology described in Section~\ref{sec:methodology_design-principles} where the dependent variable is the amount of unnecessary workload generated per day. Regression coefficients, together with corresponding Confidence Intervals (CIs), are also visually represented in the figure in Table~\ref{fig:regression_visualization}. To exemplify the interpretation of the regression coefficients, we inspect the Other Exploits group. A rule without any of the design considerations would generate $0.51$\textsuperscript{\circled{1}} alerts per day on average, and if we were to detect a characteristic directly related to the malicious behavior instead of a proxy, this workload would drop to an expected $0.51 - 0.41\textsuperscript{\circled{2}} = 0.10$ alerts per day, corresponding to a drop of $78\%$. It is possible the coefficients in a group sum to a negative number because rules in that group may typically not adhere to the design principles in a way that regression would result in a negative workload. Regression performed on the Active Scanning, Log4j and Drive-by-Compromise groups resulted in well-fitted models, as indicated by the not-rejected $H_0$ of the KS tests. Since the $H_0$ were rejected for the other groups, we can conclude that there may be other factors beyond the design principles at play that can explain the variance in unnecessary workload per day for those rules. For the Other Exploits group, we suspect the rules in that group may not have been homogenous enough with respect to the type of threat detected. Despite a lack of explained variance for the Other Exploits group, we can still trust whether design principles have positive or negative effects within that group if their p-values pass the significance test. The same holds for the Process Injection and Command \& Control groups.

We can observe that rules from the Other Exploits group leveraging a characteristic directly related to malicious behavior raise significantly fewer alerts than similar rules making use of a proxy for malicious behavior. Concretely, we observe a reduction of $78\%$\textsuperscript{\circled{2}}.
Furthermore, we observe a significant effect in the Log4j group on rules detecting successful exploitation, which cause $36\%$\textsuperscript{\circled{3}} less false positives compared to rules not distinguishing between (un)successful malicious actions. We also find a small but significant reduction in workload\textsuperscript{\circled{4}} for rules within the Drive-by-Compromise group and identify a strong reduction\textsuperscript{\circled{5}} for the Command \& Control group.
Moreover, for rules employing a form of alert throttling, we find a strong significant effect in the Log4j group, suggesting alert throttling can significantly reduce the workload generated by rules. In particular, we observe a reduction of $99\%$\textsuperscript{\circled{6}}, which corresponds to a reduction of $~67$ alerts per day if we would apply this design principle to every rule in this small group covering only a single vulnerability. We note that effect size at the level of the SOC has to be scaled up to potentially hundreds of rules, therefore significantly impacting SOC operations. These findings so far are in line with our expectations based on the principles posed in Section~\ref{sec:design-principles:formulation}.

Another significant finding is that rules using a characteristic that is generalized to match similar threats cause more workload than their peers using more specific characteristics in the Active Scanning, Log4j, and Drive-by-Compromise groups\textsuperscript{\circled{7}}. Although this design principle was intended to increase coverage, it negatively impacts workload suggesting a trade-off between coverage and specificity should be carefully considered.
In particular, we do not observe any significant effects for rules generalizing the position of the matched characteristic\textsuperscript{\circled{8}}. In further detail, we observe mixed effects when looking at the coefficients. It is possible that generalizing the position of the matched characteristic has no effect on the generated workload, considering this design principle is primarily aimed at increasing coverage.
For rules including exceptions, we also observe no significant effects in any group\textsuperscript{\circled{9}}. Potentially, significance was difficult to obtain due to the lack of rules utilizing exceptions, as reported in Table~\ref{tab:labels_distribution}. Furthermore, exceptions may be included as a coping mechanism to deal with otherwise high-noise rules causing a potential correlation between the label and the generated workload. Another potential reason for the lack of a significant effect may be that exceptions can be specific to certain environments or applications, rendering their inclusion irrelevant in other environments. Nevertheless, we can see that all other coefficients for this principle suggest a reduction of the workload generated albeit the effect is not statistically significant.

\section{Operationalization of Design Principles}
\label{sec:operationalization}
We now provide an evaluation of the prevalence of design principle violations in two common rulesets (ProofPoint's ETOPEN and ETPRO), and how the application of the proposed design principles to a subset of selected rules can impact alert generation in a SOC.

\smallskip
\noindent\textbf{Prevalence of design principles.}
\looseness=-1We devise a tool to predict whether a rule adheres to the principles using the manually labeled rules from Section~\ref{sec:methodology:effect-size-evaluation} as training data and using implementation-level features describing used options and keywords. For each principle, we train a separate eXtreme Gradient Boosting (XGBoost) classifier~\cite{xgboost, xgboost-performance} and evaluate its effectiveness by 10 repeated stratified 2-fold Cross-Validations (CV). The leveraged features and the optimization procedure using grid search are further detailed in Appendix~\ref{sec:appendix_2}. Table~\ref{tab:tool-performance-and-results} shows the CV scores for each design principle and the estimated prevalence of violations of principles in the two major rulesets (ETOPEN, ETPRO) widely employed by SOCs. The results show that the tool is capable of detecting design principles violations, with most precision scores exceeding $0.90$. The data suggests the majority of the rules focuses detection efforts on successful malicious actions and uses a generalized characteristic that is matched at a variable location, although the majority also uses proxies for detection and fails to utilize exceptions and alert throttling.

\begin{table}[tp]
  \caption{Principle Violation Detection Tool performance across the different groups and the adherence to the various design principle in the combined open and commercial rulesets acquired by the SOC according to the tool.}
\vspace{-0.15in}
  \label{tab:tool-performance-and-results}
  {\small
    \begin{tabular}{lrrrr}
      \toprule
      & \multicolumn{2}{c}{Cross-Validation} & \multicolumn{2}{c}{SOC Ruleset} \\
      Design Principle           & Precision & Recall & Count & Proportion \\
      \midrule
      Limited Proxy              & $0.81$       & $0.70$    & $42k$   & $62\%$      \\
      Successful                 & $0.99$       & $1.00$    & $18k$   & $28\%$       \\
      Exceptions                 & $0.98$       & $0.95$    & $57k$   & $85\%$      \\
      Alert Throttling           & $1.00$       & $1.00$    & $64k$   & $96\%$      \\
      Generalized Characteristic & $0.92$       & $0.62$    & $20k$   & $30\%$      \\
      Generalized Position       & $0.74$       & $0.43$    & $16k$   & $24\%$      \\
      \bottomrule
    \end{tabular}
  }
  \Description{Principle Violation Detection Tool performance across the different groups and the adherence to the various design principle in the combined open and commercial rulesets acquired by the SOC according to the tool.}
\vspace{-0.25in}
\end{table}

\smallskip
\noindent\textbf{Effect on monitoring environments.}
To assess the actionability of the design principles and to showcase the effect sizes in monitoring environments, we conduct two experiments: first, we evaluate the effect of the modified rules on alert triggers and coverage over a standard intrusion detection dataset, CIC-IDS2017~\cite{cicids2017}. Second, we evaluate the effect of deploying the improved rules in the detection environment of the collaborating SOC on alert workload.

\noindent\textit{1) Effect on event coverage.} Using the tool described above, we identify rules available to the SOC that deviate from the design principles and choose among these the rules that can detect malicious activity in the CIC-IDS2017 dataset~\cite{cicids2017}. We then chose to improve the noisiest rules, i.e. those that cause at least one FP daily in the SOC. At our request, the Tier-2 analyst who participated in the interviews (Section~\ref{sec:methodology:interviews}) is tasked to improve these rules by implementing the proposed principles.
Using the tool we establish the improved rules\footnote{Rules with {\tiny\texttt{noalert}} set are excluded to ensure groups have an equal number of rules.} together violate $34\%$ less principles.
Table~\ref{tab:improvement-coverage} shows the results on alert-triggering on the CIC-IDS2017 dataset for the original and improved rules. The results show that the improved rules detected the same attacks on CIC-IDS2017 whilst raising fewer alerts and not triggering on benign traffic or attempted attacks.\footnote{Although a more ideal coverage evaluation could use a user study to asses whether alert investigations result in similar conclusions, this is beyond the scope of the paper.} While the improved ruleset has fewer unique rules detecting the two portscan categories, an investigation of the alerts generated by the rules that only fired before the improvement reveals that these alerts correspond to scans for which no connection was successfully established, and hence no open services were discovered. These connections and the attempted bruteforce attack are not detected by the improved rules since they do not indicate successful malicious actions. Overall, we find that the improved rules trigger far fewer times while maintaining the same coverage as the original rules.

\noindent\textit{2) Effect in the SOC's environment.} We deploy the modified and original rules in the SOC and evaluate effects on triggered alerts over a period of $27$ days in production. The results indicate that the improved rules trigger $167\,245$ fewer alerts (i.e., $98\%$) than the original rules, thus significantly reducing the workload generated by these rules. We conclude that the proposed principles can help SOCs improve noisy rules to retain coverage and improve specificity, as an alternative to the typical tuning by disabling noisy rules.

\begin{table}[tp]
  \caption{Overview of coverage provided by the original and improved rules on the CIC-IDS2017 dataset.}
\vspace{-0.15in}
  \label{tab:improvement-coverage}
  {\small
    \begin{tabular}{m{1.55cm}@{}rrrrrr}
      \toprule
      & \multicolumn{3}{c}{Original}& \multicolumn{3}{c}{Improved} \\
      Label           & Rules & Alerts & \shortstack{Attack \\ Detected} & Rules & Alerts & \shortstack{Attack \\ Detected} \\
      \midrule
      Benign              & $1$       & $2$ &   \texttildelow  & $0$   & $0$ & \texttildelow      \\
      SSH-Patator                 & $1$       & $31$ & \checkmark    & $1$   & $31$ & \checkmark      \\
      Brute Force Attempted           & $1$       & $46$ &   {\footnotesize \textsuperscript{$\ast$}}\checkmark  & $0$   & $0$ &   \texttildelow    \\
      XSS       & $1$       & $320$ & \checkmark    & $1$   & $3$ & \checkmark      \\
      SQL Injection & $3$       & $17$ & \checkmark    & $3$   & $3$ & \checkmark      \\
      Infiltration Portscan              & $7$       & $105$ & \checkmark    & $2$   & $8$ & \checkmark      \\
      Portscan                 & $10$       & $148$ & {\footnotesize \textsuperscript{$\ast$}}\checkmark    & $2$    & $7$ & \checkmark       \\
      \bottomrule
    \end{tabular}
  }
    \begin{flushleft}
        {\footnotesize \quad \textsuperscript{$\ast$} \, Some labels do not (always) correspond to successful malicious actions.}
    \end{flushleft}
  \Description{Overview of coverage provided by the original and improved rules on the CIC-IDS2017 dataset.}
\vspace{-0.25in}
\end{table}

\section{Discussion}
\label{sec:discussion}

\looseness=-1In Section~\ref{sec:design-principles}, we propose six design principles to support the development of rules such that rule specificity can be maximized while preserving coverage.
We believe these principles will benefit rule developers, especially less experienced ones. Users range from junior rule developers at a SOC to malware researchers who create rules as a result of their malware investigations. These rules are often collected as part of community rulesets and may be deployed in multiple locations, with negative consequences if these were high-noise rules.

\looseness=-1The proposed design principles may also be useful to SOC engineers when tuning their rulesets to minimize FPs.
Strict adherence to the principles is nontrivial, and trade-offs may exist that compel rule engineers to deviate from a posed design principle, e.g., leveraging proxies to reduce the rule development time. An improved understanding of the effects of the principles contributes to making such trade-offs.
The simplest use case could be to review rules based on the principles and suppress rules based on how well they adhere to them.
We also envision another use case where, after review, engineers can improve new rules ingested from community and commercial rulesets or CTI feeds based on the principles, as showcased in Section~\ref{sec:operationalization}. For instance, typical rules detecting reconnaissance activity could be tailored to specific customer environments (limiting the source or destination addresses causing the rule to trigger); rules detecting exploitation attempts could be augmented to only trigger in case of evidence of successful exploitation, etc. In addition, the inclusion of exceptions, which only few rules appear to have, could greatly benefit from feedback loops as proposed in~\cite{alert-alchemy}.

\looseness=-1We make several additional observations based on our analysis of the rules applied at the collaborating SOC.
Based on the (limited) labeling of rules we did, we observe that for community and commercial rulesets, there appears to be a lack of rules distinguishing between successful and unsuccessful attacks, which could be a major improvement point.
Contrary to the relatively common practice of disabling legacy rules at SOCs~\cite{alert-alchemy}, we find that legacy rules can effectively contribute to the detection of incidents. Since no incidents within the monitoring scope remained undetected, as discussed in Section~\ref{sec:data_exploration:incident_detection}, we suspect that coverage does not require major improvements.
Furthermore, our results confirm several findings from~\cite{ruling-the-rules}, stating that a significant majority of the alerts are not related to any security incident and that there exists a large imbalance among the number of alerts generated by each rule. 
Therefore, automated approaches operating on SOC alert data, such as~\cite{Ede_Aghakhani_Spahn_Bortolameotti_Cova_Continella_Steen_Peter_Kruegel_Vigna_2022}, must be (made) robust in the presence of large imbalance and data quality issues.
Even if alert or incident data cannot be released due to a Non-Disclosure Agreement (NDA) with the SOC, we make available the tool~\footnote{\anon{\url{https://github.com/Koen1999/suricata-check}}} used to assess the adherence to the principles and the improved rules~\footnote{\anon{\url{https://github.com/Koen1999/ruling-the-unruly}}} from Section~\ref{sec:design-principles:formulation} and Section~\ref{sec:operationalization}.

\paragraph{Limitations}

\looseness=-1The generalization of the statistical inferences identified in Section~\ref{sec:design-principles:statistical:results} depends on the assumptions that (1) the SOC employs common practices; and (2) the monitored network environments are similar. Although our research leverages rule, alert, and incident data from multiple organizations over a prolonged period, it is still only reflective of a single SOC. The collaborating SOC applies several common practices also seen at other SOCs as described in related work~\cite{ruling-the-rules, alert-alchemy, Vielberth_Bohm_Fichtinger_Pernul_2020}, and we expect our findings to generalize to other SOCs monitoring similar environments. However, we deem the generalization of our findings to ICS or IoT domains a point to be addressed by future research.
Additionally, we derived our design principles from rules meant to detect three specific tactics: Reconnaissance, Initial Access, and Command \& Control. While we consider these among the most relevant for a NIDS, we cannot exclude that other specific or, less likely, even general design principles could be formulated when analyzing rules meant to detect other tactics, such as Lateral Movement.
Furthermore, we acknowledge that the identified principles may affect NIDS performance, other than specificity and coverage, which are beyond the scope of this study.

\looseness=-1The collaborating SOC strictly defines incidents as events with a security impact. As such, events of this type are relatively rare in our dataset, especially compared to some previous studies using similar data. This constitutes a challenge as fewer observations are available to make statistical inferences, limiting the possibility of investigating the effect of the principles on the number of detected incidents. On the other hand, the precise definition allows us to narrow our evaluation to events that \textit{really} matter. Furthermore, the few observations for the log4j group in Table~\ref{tab:regression_workload} suggest more data points would increase confidence for this group.
The rejection of $H_0$ in the KS-test for certain groups suggests some regression models are unsatisfactory for explaining variance. A lack of homogeneity w.r.t. what rules within these groups detect may cause variation in unnecessary workload between these rules, regardless of how the rules are written. Alternatively, variables beyond the design principles discussed in this work may further explain variance within these groups, which can be a topic for future research.
As a mitigation, we conducted the additional experiment in Section~\ref{sec:operationalization} as a sanity check, where we improved rules deviating from the principles and evaluated the effect on the specificity and coverage. This additional experiment reinforces the conclusions drawn in the prior sections.

\paragraph{Future Work}

\looseness=-1During the interviews, we discussed several factors that can influence the quality of IoC-based rules. For instance, different rules should be built based on IP addresses associated with scanning activity vs. addresses associated with C2 infrastructures. 
The unnecessary workload caused by IoC-based rules may be mitigated by quality checks performed on the IoCs.
For instance, whether an IoC is current should be monitored through its age and potentially other aspects to assess whether it should still be leveraged for detection. Considering our finding from Section~\ref{sec:data_exploration:incident_detection} that IoC-based rules contribute greatly to the incident detection, we believe future research can investigate the design of IoC-based rules more in-depth.

In addition to specificity and coverage, other requirements~\cite{habibullah-21}, such as explainability~\cite{99-false-positives}, can be important for rule design and could drive future research. Syntax and formatting, as discussed by \cite{suricata-style-guide}, can improve the maintainability of rules and the actionability of alerts. The inclusion of appropriate metadata was also indicated to be important for interpretability during the interviews. Documentation concerning the rule design, development, and revision process is currently not available for many of the inspected rules, but would be considered very beneficial to both maintainability and interpretability. 
Moreover, a lack of rule portability may be related to the noisiness of rules in some, but not all, environments.

\section{Conclusion}
\label{sec:conclusion}
In this paper, we explore data from a SOC to characterize the NIDS rules responsible for unnecessary workload.
During this process, we discover six relevant design principles, which we consolidate through interviews with experienced SOC rule designers. 
We then validate our principles by quantitatively assessing their effect on rule specificity.
We find that several of these design considerations significantly impact rule specificity. For instance, rules that leverage proxies for detection, and rules that do not employ alert throttling or do not distinguish (un)successful malicious actions, cause significantly more workload for SOC analysts.
Moreover, rules that match a generalized characteristic to detect malicious behavior, which is believed to increase coverage, significantly increase the workload as well, suggesting a trade-off must be struck between specificity and coverage. We have demonstrated that while the design principles are often violated, these principles can be applied successfully at SOCs to reduce workload whilst maintaining coverage.

\begin{acks}
The authors thank the unnamed SOC for their cooperation and for making data available for analysis, as well as the domain experts with whom we discussed the design principles. \par
\looseness=-1This publication is part of the CATRIN and INTERSECT projects (with numbers {\small NWA.1215.18.003} and {\small NWA.1160.18.301}), which is (partly) financed by the Dutch Research Council (NWO). For the purpose of Open Access, a CC-BY 4.0 public copyright license is applied to any Author Accepted Manuscript version arising from this submission.
\end{acks}

% \printbibliography
\bibliographystyle{ACM-Reference-Format}
\bibliography{main}

\appendix

\section{Suricata rule syntax}
\label{sec:appendix_0}
\begin{figure}[H]
    \includegraphics[width=1.0\linewidth]{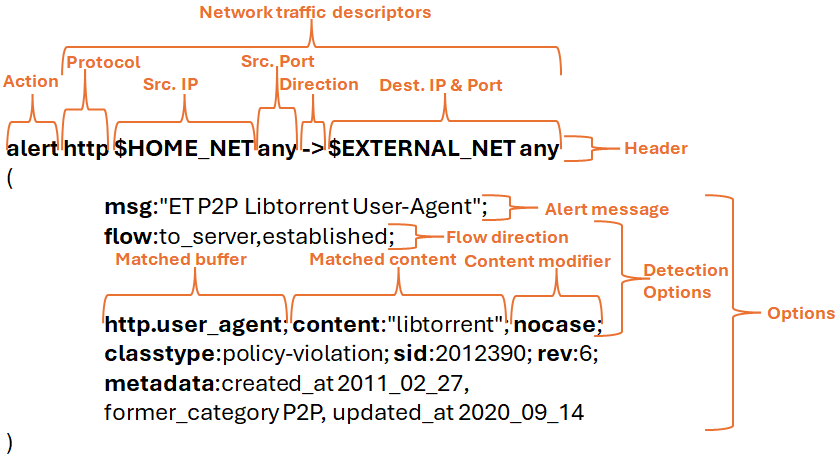}
\vspace{-0.25in}
    \caption{Rule syntax used by Suricata NIDS.}
    \label{fig:suricata-example}
    \Description{Rule syntax used by Suricata NIDS.}
    \vspace{-0.15in}
\end{figure}

\section{Dataset key statistics}
\label{sec:appendix_1.25}
\begin{table}[hbt]
    \caption{Overview of key statistics (mean and standard deviation) of the collected dataset, covering rules and alerts.}
\vspace{-0.15in}
    \label{tab:data_overview}
    {\small
    \begin{tabular}{lr}
        \toprule
        Daily Uniquely Triggered Rules & $\simeq41\pm18$ \\
        Daily Uniquely Observed Techniques & $\simeq8\pm2$ \\
        Daily Uniquely Observed CVEs & $\simeq3\pm4$ \\
        Daily Alerts & $46\,309\pm75\,660$ \\
        Minimum Daily Attackers & $26\pm10$ \\
        Maximum Daily Attackers & $3\,775\pm936$ \\
        Minimum Daily Targets & $8\pm7$ \\
        Maximum Daily Targets & $1286\pm39$ \\
        Unnecessary Workload per Rule & $21\pm282$ \\
        \bottomrule
    \end{tabular}
    }
    \Description{Overview of key statistics (mean and standard deviation) of the collected dataset, covering rules and alerts.}
    \vspace{-0.15in}
\end{table}

\section{Additional examples of Intrusion Detection Rules}
\label{sec:appendix_1}
\begin{figure}[H]
\begin{lstlisting}
alert http $HOME_NET any -> $EXTERNAL_NET any (
msg:"ET MALWARE Terse alphanumeric executable downloader high likelihood of being hostile";
flow:established,to_server;
http.uri; content:"/"; content:".exe"; distance:1; within:8; endswith; pcre:"/\/[A-Z]?[a-z]{1,3}[0-9]?\.exe$/";
http.header; content:!"koggames"; http.host; content:!"download.bitdefender.com"; endswith; content:!".appspot.com"; endswith; content:!"kaspersky.com"; endswith; content:!".sophosxl.net"; endswith;
http.header_names; content:!"Referer"; nocase;)
\end{lstlisting}
\vspace{-0.15in}
    \caption{Shortened rule detecting executable downloads with numerous exceptions for common anti-virus software (sid: $2019714$)}
    \label{fig:executable-download-exceptions}
    \Description{Shortened rule detecting executable downloads with numerous exceptions for common anti-virus software (sid: $2019714$)}
\vspace{-0.15in}
\end{figure}
\begin{figure}[H]
\begin{lstlisting}
alert http any any -> $HOME_NET any (
msg:"ET EXPLOIT D-Link DSL-2750B - OS Command Injection";
flow:established,to_server;
http.uri; content:"/login.cgi?cli="; pcre:"/^[ a-zA-Z0-9+_]*[\x27\x3b]/Ri";)
\end{lstlisting}
\vspace{-0.15in}
    \caption{Shortened rule detecting command injection using a fixed location for the HTTP GET query parameter name (sid: $2025756$)}
    \label{fig:command-injection-generalized-location}
    \Description{Shortened rule detecting command injection using a fixed location for the HTTP GET query parameter name (sid: $2025756$)}
\vspace{-0.15in}
\end{figure}
\begin{figure}[H]
\begin{lstlisting}
alert http $EXTERNAL_NET any -> $HOME_NET any (
msg:"ET SCAN OpenVASVT RCE Test String in HTTP Request Inbound";
flow:established,to_server;
content:"T3BlblZBU1ZUIFJDRSBUZXN0";
threshold:type limit, track by_src, count 1, seconds 60;)
\end{lstlisting}
\vspace{-0.15in}
    \caption{Shortened rule detecting inbound OpenVAS User-Agent (sid: $2033101$)}
    \label{fig:rule-openvas-better}
    \Description{Shortened rule detecting inbound OpenVAS User-Agent (sid: $2033101$)}
\vspace{-0.15in}
\end{figure}
\begin{figure}[H]
\begin{lstlisting}
alert http any any -> $HTTP_SERVERS any (
msg:"ET WEB_SERVER ColdFusion adminapi access";
flow:established,to_server;
http.method; content:"GET"; nocase;
http.uri; content:"/CFIDE/adminapi";)
\end{lstlisting}
\vspace{-0.15in}
    \caption{Shortened rule detecting Coldfusion unauthorized adminapi access attempts (sid: $2016183$)}
    \label{fig:rule-coldfusion-2}
    \Description{Shortened rule detecting Coldfusion unauthorized adminapi access attempts (sid: $2016183$)}
\vspace{-0.15in}
\end{figure}

\section{Training procedure of Principle Adherence Detection Tool}
\label{sec:appendix_2}
The tool's goal is to predict whether a rule adheres to the principles after being trained on the $182$ manually labeled rules from Section~\ref{sec:methodology:effect-size-evaluation} and using implementation-level features describing used detection options and keywords.
One group of features are integers describing how often (and hence also whether) an option is used in a rule. This feature was computed for the {\small\texttt{content}}, {\small\texttt{depth}}, {\small\texttt{http.uri}}, {\small\texttt{http.method}}, {\small\texttt{urilen}}, {\small\texttt{startswith}}, {\small\texttt{pcre}}, and {\small\texttt{bsize}} options. Additionally, we devise an additional feature by counting the number of negated matches performed. We also devise features that describe whether the source and destination IP addresses specified by the rule use address groups like {\small\texttt{\$HOME\_NET}}, {\small\texttt{\$HTTP\_SERVERS}}, {\small\texttt{\$EXTERNAL\_NET}}, and {\small\texttt{any}}. We also include features describing the values set for the {\small\texttt{threshold.type}} and {\small\texttt{threshold.count}} options. The last group of features describes whether the {\small\texttt{flow}} options {\small\texttt{to\_server}} and {\small\texttt{to\_client}}, or their equivalents, are set by the rule. \par
For each principle, an eXtreme Gradient Boosting (XGBoost) classifier~\cite{xgboost} is trained separately and evaluated by 10 repeated stratified 2-fold Cross-Validations (CV). XGBoost is known to perform well on a various tasks and can outperform deep neural networks with more parameters~\cite{xgboost-performance}. Using grid search on the grid\footnote{Where unspecified the default parameters of the XGBoost Python implementation are used.} shown in Table~\ref{tab:gridsearch}, hyperparameters are optimized for a weighted variant of the F1-score where precision was assigned a $10$ times greater weight than recall. The resulting weighted F1-scores ranged from $0.72$ to $1.00$.
\begin{table}[hbt]
    \caption{Grid search optimization parameters for the XGBoost algorithm.}
\vspace{-0.15in}
    \label{tab:gridsearch}
    {\small
    \begin{tabular}{lr}
        \toprule
        Number of decision trees & $1000$ \\
        Loss function & Logistic \\
        Eta (learning rate) & $[0.01, 0.1, 0.3]$ \\
        Sample sampling rate & $1.0$ \\
        Feature sampling rate & $[0.25, 0.5, 0.75, 1.0]$ \\
        Sample weight scaling & $[0.1, 0.25, 0.5, 1.0, 2.0, 4.0, 10.0]$ \\
        Maximum tree depth & $[1, 3]$ \\
        Minimum child weight & $[1]$ \\
        Gamma & $[0, 0.1]$ \\
        Lambda (L2 regularization) & $[0, 0.01, 0.1]$ \\
        Alpha (L1 regularization) & $[0, 0.01, 0.1]$ \\
        \bottomrule
    \end{tabular}
    }
    \Description{Grid search optimization parameters for the XGBoost algorithm.}
    \vspace{-0.15in}
\end{table}

\end{document}